\documentclass[11pt,a4paper]{article}
\usepackage{jcappub}
\usepackage{graphicx}
\usepackage{amsmath}
\usepackage{epsfig}
\usepackage{bm}
\usepackage{graphicx}
\usepackage{epsfig}
\usepackage{bm}
\usepackage{amsmath, amsfonts, amssymb, amstext}
\newcommand{\be}{\begin{equation}}
\newcommand{\ee}{\end{equation}}
\newcommand{\beq}{\begin{eqnarray}}
\newcommand{\eeq}{\end{eqnarray}}

\title{Photon paths in the hyperbolic topological black hole spacetime of conformal Weyl gravity}
\author[a]{Graeme Candlish}
\author[b]{ Marco Olivares}
\author[c]{Constanza Osses}
\author[a]{ J. R. Villanueva\note{Corresponding author.}}
\affiliation[a]{\it Instituto de F\'{\i}sica y Astronom\'{\i}a, Universidad de Valpara\'{\i}so, Avenida Gran Breta\~na  1111, Casilla 5030, Playa Ancha, Valpara\'{\i}so, Chile}
\affiliation[b]{\it Facultad de Ingenier\'ia y Ciencias, Universidad
Diego Portales, Avenida Ej\'ercito Libertador 441, Casilla 298--V,
Santiago, Chile.}
 \affiliation[c]{\it Instituto de F\'{\i}sica, Pontificia Universidad Cat\'olica de Valpara\'{\i}so, Avenida Universidad 330, Curauma, Valpara\'{\i}so, Chile}
 \emailAdd{graeme.candlish@uv.cl}
\emailAdd{marco.olivaresr@mail.udp.cl}
 \emailAdd{constanza.osses.g@mail.pucv.cl}
 \emailAdd{jose.villanueva@uv.cl}

\date{\today}

\abstract{In this work we find analytical solutions to the null geodesics in the hyperbolic topological black hole spacetime of conformal Weyl gravity in an invariant $2$-plane given by the orbits of an azimuthal Killing vector. Exact expressions for the (non-compact) horizons are found, which depend on the cosmological constant and the coupling constants of conformal Weyl gravity. The angular motion is examined qualitatively by means of an effective potential; quantitatively, the equation of motion is solved in terms of the $\wp$-Weierstra{\ss} elliptic function. Thus, we find the deflection angle for photons without using any approximation, which is a novel result for this topology.}


\begin{document}
\maketitle


\section{Introduction: Conformal Weyl gravity and null geodesics}\label{introduction}
 Conformal Weyl gravity (CWG) \cite{W17,W18A,W18B,B21} corresponds to a fourth order theory in which the spacetime geometry is invariant under the local change
	\begin{equation}\label{b1}
	g_{\mu \nu}(x)\rightarrow \Omega^2(x)\,g_{\mu \nu}(x),
	\end{equation}
	where $\Omega(x)$ is any smooth strictly positive function. The restrictive conformal invariance then leads to the unique action
	\begin{equation}\label{b2}
	I_W=-\alpha\,\int d^4x\,\sqrt{-g}\,C_{\lambda \mu \nu \kappa} C^{\lambda \mu \nu \kappa}
	=-2 \alpha\,\int d^4x\,\sqrt{-g} \left[R_{\mu \kappa} R^{\mu \kappa}-\frac{1}{3} R^2 \right],
	\end{equation}
	where $\alpha$ is the dimensionless gravitational coupling constant, and $C_{\lambda \mu \nu \kappa}$ is the conformal Weyl tensor given by
	\begin{equation}\label{b3}
	C_{\lambda \mu \nu \kappa}=R_{\lambda \mu \nu \kappa}-\frac{1}{2}(g_{\lambda \nu} R_{\mu \kappa}-g_{\lambda \kappa} R_{\mu \nu}-g_{\mu \nu} R_{\lambda \kappa}+g_{\mu \kappa} R_{\lambda \nu})+\frac{1}{6} R (g_{\lambda \nu} g_{\mu \kappa}-g_{\lambda \kappa} g_{\mu \nu}).
	\end{equation}
	Note that the action (\ref{b2}) naturally appears as the difference between a Weyl--squared term and the Gauss--Bonnet invariant in 3+1 dimensions. Although the Gauss--Bonnet term is a topological invariant in that case, its addition is far from trivial, since one can prove that this is equivalent to the renormalization program in the context of AdS / CFT \cite{Miskovic:2014zja}.

	The gravitational field equations associated to the action (\ref{b2}) are given by
	\begin{equation}\label{b4}
	\sqrt{-g} \, g_{\mu \alpha}\,g_{\nu \beta} \frac{\delta I_W}{\delta g_{\alpha \beta}}=-2\,\alpha W_{\mu \nu}=-\frac{1}{2}T_{\mu \nu},
	\end{equation}
	where $T_{\mu \nu}$ is the stress--energy tensor, and $W_{\mu \nu}$ is the Bach tensor given by
	\begin{equation}\label{b5}
	W_{\mu \nu}=2C^{\alpha \, \, \, \, \beta}_{\, \,  \mu \nu \, \, ;\beta \alpha}+C^{\alpha \, \, \, \, \beta}_{\, \,  \mu \nu} R_{\alpha \beta}.
	\end{equation}
	
	An exact vacuum solution to the field equations (\ref{b4}) is given, up to a conformal factor and in the usual Schwarzschild coordinates ($\tilde{t}, \tilde{r}, \Theta, \Phi$), by the static spherically symmetric metric \cite{riegert84,MK89,MK91,MK91B,MK94}
	\begin{equation}
	{\rm d}\tilde{s}^{2}=-B(\tilde{r})\,{\rm d}\tilde{t}^{2}+\frac{{\rm d}\tilde{r}^{2}}{B(\tilde{r})}+
	\tilde{r}^{2}\,d\Omega^2, \label{metrweyl}
	\end{equation}
	where $d\Omega^2={\rm d}\Theta^{2}+\sin^{2}\Theta\, {\rm d}\Phi^{2}$ is the standard metric on the 2--sphere, and the lapse function is
	\begin{equation}B(\tilde{r})=1-\frac{(2-3\,M\,\tilde{\gamma})M}{\tilde{r}}-3\,M\,\tilde{\gamma}+\tilde{\gamma} \tilde{r}
	- k \tilde{r}^{2}.
	\label{lpasweyl}
	\end{equation}
	
	Here $M$ is the central mass, $k=\Lambda/3$ is the cosmological constant and $\tilde{\gamma}$ is the Weyl parameter, which can be related to the interior dynamics of the static source of interest and the properties of the background geometry. A great deal of investigation has taken place in conformal gravity.
	For example, the treatment of bending of light from various points of view can be found in \cite{edery98,PI04,PII04,Bha09,sultana10,sultana12,vo13,FGG07}, while the gravitomagnetic and Sagnac effects are investiged in \cite{said13} and \cite{sultana14}, respectively. Black hole solutions with other properties such as electric charge, angular momentum and non-trivial topology are found in \cite{MK91A,klemm98,said12,Lu12,pf12}, and also a general class of wormhole geometries in CWG is analyzed in \cite{lobo}.
	On the other hand, some cosmological  implications have been studied in the literature, see for example \cite{Knox,Diaferio,Diaferio08,Mann12,Varieschi}, whereas some studies investigating finite mass distributions such as stars may be found in \cite{Brihaye,Barabash}. An study of the motion of massless particles in the exterior space-time of a toroidal topological black hole can be found in \cite{mrfloyd}.


Study of the motion of test particles can be obtained using the standard Lagrange procedure \cite{chandra,COV05,shutz,OSLV11,jaklitsch,VSOC13,Halilsoy:zva,LSV11,wald}, which makes it possible to associate a Lagrangian $\mathcal{L}$ with the metric and then obtain the equation of motion from Lagrange's equations
\be 
\dot{\Pi}_{q} - \frac{\partial \mathcal{L}}{\partial q} = 0,
\label{lageq} \ee
where $\Pi_{q} = \partial \mathcal{L}/\partial \dot{q}$
are the conjugate momenta to the coordinate $q$,
and the dot denotes a derivative with respect to
the affine parameter $\tau$ along the geodesic.
In the next sections, following the procedure performed by Klemm \cite{klemm98} and taking into account that the metric (\ref{metrweyl}), we perform an analytical continuation to obtain a non-trivial topology associated with black holes in CWG, and then we study the null geodesics on these manifolds.

Finally, as we can see from the metric (\ref{metrweyl})--(\ref{lpasweyl}), in this work we will use geometrized units ($G=1$ and $c=1$), so the parameters have the typical values  $k\approx 4.1\times 10^{-20}\,pc^{-2}$ and $\tilde{\gamma}\approx 3.1\times10^{-10}\,pc^{-1}$.
\section{Hyperbolic topological black hole spacetime in conformal Weyl gravity}

In order to obtain a class of
topological black hole solutions, we perform
the analytical continuation suggested 
by \cite{klemm98}:
\begin{equation}
\label{contanal1}
{\rm d}\tilde{s}\rightarrow \frac{{\rm d}s}{M}, \,\,\,k\rightarrow M^2\,k,\,\,\, \tilde{\gamma} \rightarrow i\,M \gamma,\,\,
M \rightarrow \frac{-i}{M},\,\,
\tilde{t}\rightarrow \frac{i\,t}{M},\,\,\,
\tilde{r}\rightarrow \frac{i\,r}{M},\,\,\,
\Phi\rightarrow \phi,\,\,\, 
\Theta\rightarrow i\,\theta,
\end{equation}
so  Eqs. (\ref{metrweyl})--(\ref{lpasweyl}) yield to
\begin{eqnarray}
&&{\rm d}s^{2}=-B(r)\,{\rm d}t^{2}+\frac{{\rm d}r^{2}}{B(r)}+r^{2}({\rm d}\theta^{2}+\sinh^{2}\theta\,
{\rm d}\phi^{2}), \label{metrhyp}\\
&& B(r)=-1-\frac{(2-3\gamma)}{r}+3\gamma+\gamma r
- k r^{2},
\label{lapshyp}
\end{eqnarray}
and the Lagrangian reads
\begin{equation}
\label{lagrangian}
\mathcal{L}=\frac{1}{2}B(r)\,\dot{t}^2+\frac{\dot{r}^2}{2 B(r)}+\frac{r^2}{2}(\dot{\theta}^{2}+\sinh^{2}\theta\,\dot{\phi}^{2}).
\end{equation}
Clearly, the metric induced
on the spacelike  surface of constants $t$ and $r$ is the standard
metric of hyperbolic 2-space $\mathbb{H}^2$, 
${\rm d}\sigma^{2}=r^{2}({\rm d}\theta^{2}+\sinh^{2}\theta\,
{\rm d}\phi^{2})$, which has a constant negative curvature and is not compact.
 In \cite{klemm98} the	$(\theta,\phi)$--sector is compactified by acting with an appropriate discrete subgroup $G$ of the isometry group $SO(2; 1)$ of $\mathbb{H}^2$, to form the compact quotient space $\mathbb{H}^2/G$. Upon demanding an orientable manifold, this space becomes a Riemann surface $S_g$ of genus $g > 1$, and the topology of the four-dimensional manifold is $\mathbb{R}^2\times S_g$ (the details of the compactification procedure are given in \cite{balasz}). In this work we will not apply this compactification procedure, but will instead consider the non-compact hyperbolic 2-space $\mathbb{H}^2$. The spatial slices of the event horizon in this solution are therefore non-compact hyperbolic spaces, and the solution may be considered an exotic type of black membrane.

 As discussed in \cite{klemm98}, additional parameter choices may be made to construct black hole spacetimes with non-trivial topology: i) Setting $\gamma=0$ and $k=\Lambda/3 < 0$, where  $\Lambda$ is the cosmological constant, one recovers
the uncharged, static topological black hole solutions in AdS gravity. ii) We may set $k=0$, yielding a solution which is not asymptotically AdS. iii) For $k>0$,
$\gamma>0$  and $\gamma_e<\gamma< 2/3$ the spacetime has a black hole interpretation, $\gamma=\gamma_e$ yielding an extreme
black hole, where $\gamma_e^2(1-\gamma_e)/(2-3\gamma_e)^2=k$.
On the other hand, it is not hard to show that the scalar curvature $R$ diverges at $r=0$ so we have a curvature singularity at the origin. In addition, $R$ vanishes for $r\rightarrow\infty$ if $k=0$, although the manifold is not Ricci-flat at infinity and therefore is not asymptotically Minkowski space. For this value of $k$ the manifold is globally hyperbolic, which is not the case for AdS black holes, see \cite{klemm98}.

\begin{figure}[tbp]
 \centering
   \includegraphics[width=90mm]{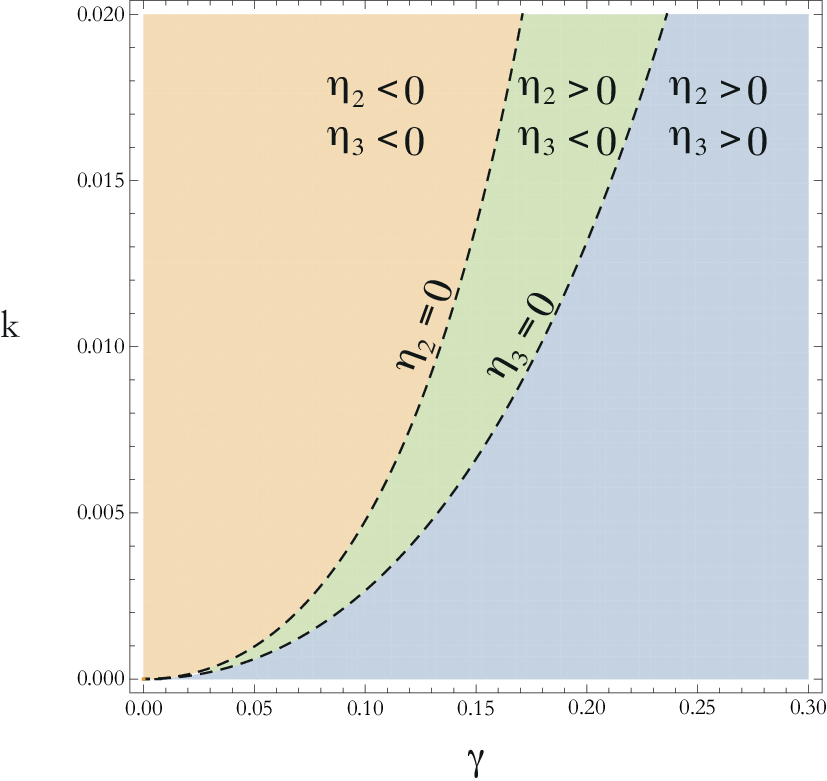}
 \caption{\label{graf1} This region plot shows the sign of the cubic parameters of the polynomial (\ref{polycanon}), $\eta_2$ and $\eta_3$, on the $k$-$\gamma$ plane. The region in which $\eta_2<0$ and $\eta_3<0$ does not allow a real solution to the equation $p_3(x)=0$.}
 \end{figure}

From the hyperbolic lapse function (\ref{lapshyp}), it is possible to construct the cubic polynomial
\begin{equation}
p_3(r)=-\frac{r B(r)}{k}=r^{3}-\frac{\gamma}{k}r^2 +\frac{(1-3\gamma)}{k}r+\frac{(2-3\gamma)}{k},
\label{cubpoly}
\end{equation}
which gives us the locations of the horizons (if there are any). 
In fact, defining $r_w=\gamma/3k$ and then
making the change of variable:
\begin{equation}
r=x+r_w,
\label{cv1}
\end{equation}
we write the cubic polynomial in its
canonical form
\begin{equation}
p_3(x)=x^3-\eta_2\,x-\eta_3,
\label{polycanon}
\end{equation}
where the coefficients are given by
\begin{eqnarray}
&&\eta_2=3\,r_w^2+9\, r_w-\frac{1}{k},\label{coef1}\\
&&\eta_3=2\,r_w^3+9\,r_w^2+
\left(9-\frac{1}{k}\right)\,r_w-\frac{2}{k}.
\label{coef2}
\end{eqnarray}
Obviously, the nature of the coefficients depends on the combination of the pair $(\gamma, k)$. 
In Fig. \ref{graf1} three different regions appear with the possible choice of the sign of $\eta_2$ and $\eta_3$ on the $k$-$\gamma$ plane. 
In the region between $\gamma=0$ and $\eta_2=0$, the roots of the polynomial $p_3(x)$ satisfy the equation $x^3+|\eta_2|\,x+|\eta_3|=0$, so there is no real solution for $x>0$, and thus we dismiss this case.
On the other hand, one may obtain information about the roots from the cubic 
discriminant $\Delta_c=27 \eta_3^2-4\eta_2^3$.
Thus, if
$\Delta_c<0$, there is one real negative root and a
complex pair of roots, representing a naked singularity; if $\Delta_c>0$, there are three different real roots, two positive and one negative, which looks similar to the SdS spacetime with an event and cosmological horizons.
Finally, if $\Delta_c=0$, there are three real roots,
one negative and a degenerate positive root, representing the extreme case. 
Thus, for example, if we assume that $\gamma$ is small,
then the coefficient $\eta_3$ is negative,
and therefore the discriminant of the polynomial
is positive, $\Delta_c>0$.

Denoting
\begin{equation}
R=\sqrt{\frac{|\eta_2|}{3}},\quad
\varphi=\frac{1}{3}\arccos\left(\frac{\eta_3}{R^3}\right),
\end{equation}
we can find the expression for the event horizon, $r_+$,
the cosmological horizon, $r_{++}$, and the negative root
(without physical meaning), $r_n$. They take the form
\begin{eqnarray}
&&r_+= r_w+\frac{R}{2}(\cos \varphi-\sqrt{3}\sin\varphi),\label{eh}\\
&&r_{++}= r_w+R\cos\varphi,\label{w.7}\\
&&r_n= r_w-\frac{R}{2}(\cos \varphi+\sqrt{3}\sin\varphi).\label{ch}
\end{eqnarray}

Conversely, the (non-compact) hyperbolic metric (\ref{metrhyp})
admits the following Killing vector fields \cite{lightman}:

\begin{itemize}
\item the {\it time-like Killing vector} $\xi=\partial_t$ is related to the {\rm 
stationarity} of the metric. The conserved quantity is given by
\begin{equation}
\label{energy}
g_{\mu \nu}\,\xi^{\mu}\,u^{\nu}=-B(r)\,\dot{t}=-\sqrt{E}
\end{equation}
where $E$ is a constant of the motion that cannot be
associated with the total energy of the test particle
because this metric is not asymptotically flat.
\item the most general \textit{space-like Killing vector}
is given by
\begin{equation}
\vec{\xi}= \left(A\,\cos \phi+B\,\sin\phi \right)\,\partial_{\theta}
+\left[C-\coth\theta\,\left(A\,\cos \phi+B\,\sin\phi \right)\right]\,\partial_{\phi},
\end{equation}
where $A$, $B$ and $C$ are arbitrary constants.
It is easy to see that this is a linear combination of the three
Killing vectors
$$\chi_1=\partial_{\phi},\quad \chi_2= -\coth \theta\,\sin \phi\,
\partial_{\phi}+\cos \phi\,\partial_{\theta},\quad
\chi_3=\coth \theta\,\cos \phi\,
\partial_{\phi}+\sin \phi\,\partial_{\theta},$$ which are 
the angular momentum operators for this spacetime. The conserved quantities
are given by
\begin{eqnarray}\label{cca0}
g_{\alpha \beta }\,\chi_{1}^{\alpha}\,u^{\beta}
&=&r^{2}\,\sinh^2 \theta\,\dot{\phi}
=L_1, \\ \label{cca1}
g_{\alpha \beta }\,\chi_{2}^{\alpha}\,u^{\beta}
&=&r^{2}\,(\cos\phi\,\dot{\theta}-\sinh\theta\,\cosh \theta \sin\phi\,\dot{\phi})
=L_2, \\ \label{cca2}
g_{\alpha \beta }\,\chi_{3}^{\alpha}\,u^{\beta}
&=&
r^{2}\,(\sin\phi\,\dot{\theta}+\sinh\theta\,\cosh \theta\, \cos\phi\,\dot{\phi})
=L_3,
\end{eqnarray}
where $L_1$, $L_2$ and $L_3$ are constants associated with the angular momentum
of the particles. Thus, from Eqs. (\ref{cca0}--\ref{cca2}), it is easy to demonstrate that
\begin{equation}
\label{condmom}\frac{J^2}{r^4}\equiv \frac{(L^2_2+L^2_3)-L^2_1}{r^4}=\dot{\theta}^2+\sinh^2 \theta \dot{\phi}^2.
\end{equation}
We now focus our attention, without loss of generality, on the invariant $2$-plane $\theta={\rm arcsinh} (1)$, so $\dot{\theta}=0$, and therefore Eq. (\ref{condmom}) reduces to
\begin{equation}
\label{angeqmot}
\dot{\phi}=\frac{J}{r^2}.
\end{equation}
\end{itemize}

\begin{figure}[tbp]
 \centering
   \includegraphics[width=150mm]{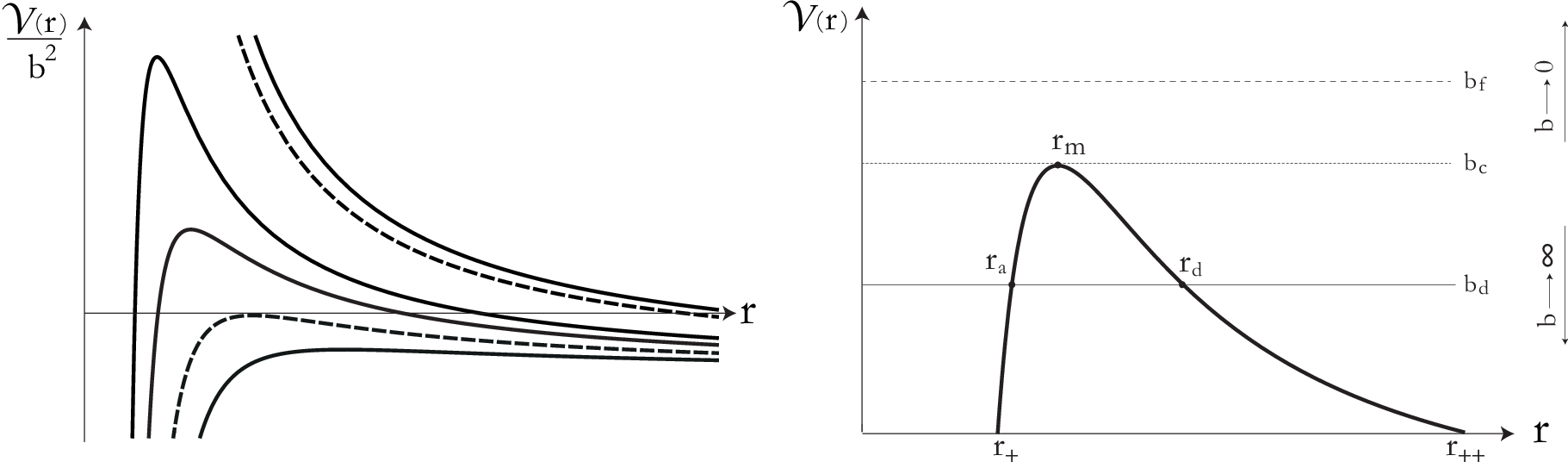}
 \caption{Left panel: Evolution of the conformal effective potential,
 per unit of $b^2$,
  for photons in Weyl's gravity as a function of the
  Weyl parameter $\gamma$ with a fixed value of the cosmological parameter $k=0.05$. The 
upper dashed line corresponds to the curve with $\gamma=\gamma_c$
while the lower dashed line corresponds to the extreme topological black hole, for which the relation $V_m=0$ is satisfied.
Right panel:   Typical effective potential with three general
regions for the motion of photons: deflection zone, in which the impact parameter poses a value between $[b_c<b\equiv b_d<\infty]$,
asymptotic zone, in which the impact parameter has the value $b=b_c$,
and a capture zone in which the impact parameter has a value
between $[0<b\equiv b_f<b_c]$.}
  \label{f2} 
 \end{figure}

Using Eqs. (\ref{energy}) and (\ref{angeqmot}), and taking into account that the Lagrangian associated with the motion of photons is null, $\mathcal{L}=0$, it is easy to obtain the radial equation of motion
\begin{equation}
\dot{r}^{2}=E-V(r),
\label{lagrhyp}
\end{equation}
where $V(r)$ is the {effective potential}
which satisfies the relation
\begin{equation}
\label{effpot}
\frac{V(r)}{E}\equiv \mathcal{V}(r)=b^2\,\frac{B(r)}{r^2},
\end{equation}
and $b=J/\sqrt{E}$
is the impact
parameter. This last point is relevant for the radial motion, since $V(r)\propto b^2=0$; the familiar and well studied relation $\dot{r}=\pm\sqrt{E}$ is recovered.
In its essence, the motion presents the same features known in other spacetimes (see for example \cite{chandra,COV05,shutz}).
On the other hand, by combining Eqs. (\ref{angeqmot}) and (\ref{lagrhyp}), we obtain the radial--polar equation
\begin{equation}
\left(\frac{dr}{d\phi}\right)^{2}= \frac{r^4}{b^2}\left[1-
\mathcal{V}(r)\right].
\label{radialpolareq}
\end{equation}
In Fig.\ref{f2} we have plotted the conformal effective potential (per unit of $b^2$) as a function of the radial coordinate, for a fixed value of the cosmological parameter $k=0.05$.
The evolution of the curves was made in terms of the Weyl parameter $\gamma$. Thus, as with Einstein's black hole, the position of the maximum $r_m$ is independent of $k$
and is thus only a function of $\gamma$:
\begin{equation}
\label{rmax}
r_m=3\left(\frac{\gamma_c}{\gamma}-1\right),
\end{equation}
where $\gamma_c=2/3$. At this point the impact parameter assumes the critical value
\begin{equation}
b_c=\frac{1}{\sqrt{k_{ext}-k}},
\end{equation}
where $k_{ext}=(1-\gamma)/r_m^2$.
In the same figure, the upper dashed line represents the effective potential for $\gamma_c$, which implies that $r_m=0$ and $V(r_m)\equiv V_m\rightarrow \infty$,
whereas on the lower dashed line the relation
$k=k_{ext}$ is satisfied,
which represents the {\it extreme case} for which
$V_m=0$. Note that the latter condition implies that 
$b_c\rightarrow \infty$.
With this in mind, we focus our attention on the family of 
potentials that lie between these two well defined curves.
The next subsection is devoted to analyzing
the effective potential qualitatively in terms of the impact parameter.


\subsection{Allowed orbits}
As we have mentioned, under the assumption
that $\gamma< \gamma_c$ and $k<k_{ext}$, massless particles can perform
different kinds of orbits, which depend on their impact parameter.
In this way, based on the classification of $b$ shown in the left panel of Figure \ref{f2},
we present a brief qualitative description of the angular motions permitted for photons in the chosen invariant $2$-plane of the non-compact hyperbolic topological black hole spacetimes in conformal Weyl gravity.
\begin{itemize}
  \item \emph{Unbounded trajectories (capture zone)}:
  If $0<b\equiv b_f <b_{c}$, the incident photons fall inexorably
  to one of the two horizons, depending on the initial conditions placed
  on their velocity.
  The cross section,
  $\sigma$, in this geometry is \cite{wald}
  \begin{equation}\label{mr51}
    \sigma=\pi\,b_c^2=\frac{\pi}{k_{ext}-k}.
  \end{equation}
  \item \emph{Critical trajectories}:
  If $b=b_{c}$, photons can stay in  the unstable
  inner circular orbit of radius  $r_{m}$.
  Therefore, the photons that arrive from the initial distance
  $r_i$ ($r_+ < r_i< r_m$, or $r_m< r_i<r_{++}$)
  can asymptotically fall to a circle of radius $r_{m}$.
  The proper period in such an orbit is given by
  \begin{equation}\label{p1}
  T_{\tau}=\frac{2\,\pi\,r_m^2}{J}=\frac{18\,\pi}{J}\left(
  \frac{\gamma_c}{\gamma}-1\right)^2,
  \end{equation}
  which is independent of $k$,
  whereas the coordinate period depends on $k$ and $\gamma$ as
  \begin{equation}\label{p2}
  T_t=2\,\pi\,b_c=\frac{2\,\pi}{\sqrt{k_{ext}-k}}.
  \end{equation}
  \item \emph{Deflection Zone}. If $b_{c} <b=b_d <\infty $, the
  photons come from a finite  distance $r_i$ ($r_+ < r_i< r_m$ for orbits of the first kind or $r_m< r_i<r_{++}$ for orbits of the second kind)
  to a turning point $r_{t}$ (which is the solution to the equation $\mathcal{V}(r_t)=1$), and then return to one of the two horizons. Therefore, by defining \begin{equation}\label{defRw} R_W=\frac{\gamma\,\mathcal{B}^2 }{3},\quad \bar{R}=\sqrt{\frac{\varrho_2}{3}},\quad \zeta=\frac{1}{3}\arccos\sqrt{\frac{27 \varrho_3^2}{\varrho_2^3}},\end{equation} where the cubic coefficients are given by
  \begin{equation}\label{mr54}
  \varrho_2=3\,\gamma_c^2\,\mathcal{B}^2
  \left[\gamma^2\,\mathcal{B}^2+3
  \left(\frac{3-r_m}{3+r_m}\right)\right],
  \end{equation}
  and
  \begin{equation}\label{mr55}
  \varrho_3=\gamma_c^2\,\mathcal{B}^2
  \left[\gamma_c\,\gamma^3\,\mathcal{B}^4
  +\,\frac{6\,(3-r_m)}{(3+r_m)^2}\,\mathcal{B}^2-
  \frac{18\,r_m}{3+r_m}
  \right],
  \end{equation}
  and $\mathcal{B}$ is the {\it anomalous impact parameter}
  given by
  \begin{equation}\label{anom}
  \mathcal{B}=\frac{b}{\sqrt{1+k\,b^2}},
  \end{equation} these solutions are given explicitly through the following equations: \begin{eqnarray}\label{defr1}
  	r_1&=&R_W-\frac{\bar{R}}{2}\left(\sqrt{3}\sin \zeta + \cos \zeta \right),\\ \label{distapelio} r_a&=&R_W+\frac{\bar{R}}{2}\left(\sqrt{3}\sin \zeta - \cos \zeta \right),\\ \label{distaperi} r_d&=&R_W+\bar{R}\,\cos \zeta.
  \end{eqnarray} Notice that if the condition  $\varrho_2^3>27 \varrho_3^2$ is assumed, it is not difficult to show that the preceding solutions satisfy $r_1<0<r_a<r_d$. Also note that, as has been pointed out by Cruz et al. \cite{COV05}, since $k$ is very small, only when
 $b\sim k^{-1/2}$ is the net effect of $\mathcal{B}$ relevant in (\ref{anom}). Also, if the product
 $k\,b^2 \gg 1$ (or $b\rightarrow\infty$), then $\mathcal{B} \sim k^{-1/2}$
 in such a way that $\varrho_2 \rightarrow \eta_2$ and
 $\varrho_3 \rightarrow \eta_3$ (cf. Eqs. (\ref{coef1})--(\ref{mr54}) and 
 Eqs. (\ref{coef2})--(\ref{mr55})). Therefore, 
 as expected, the identities $r_a(\infty, \gamma, k)=r_{+}$ and  $r_d(\infty, \gamma, k)=r_{++}$
 can be proven in a few steps \cite{vo13}.

\end{itemize}

\begin{figure}[tbp]
	\centering
	\includegraphics[width=150mm]{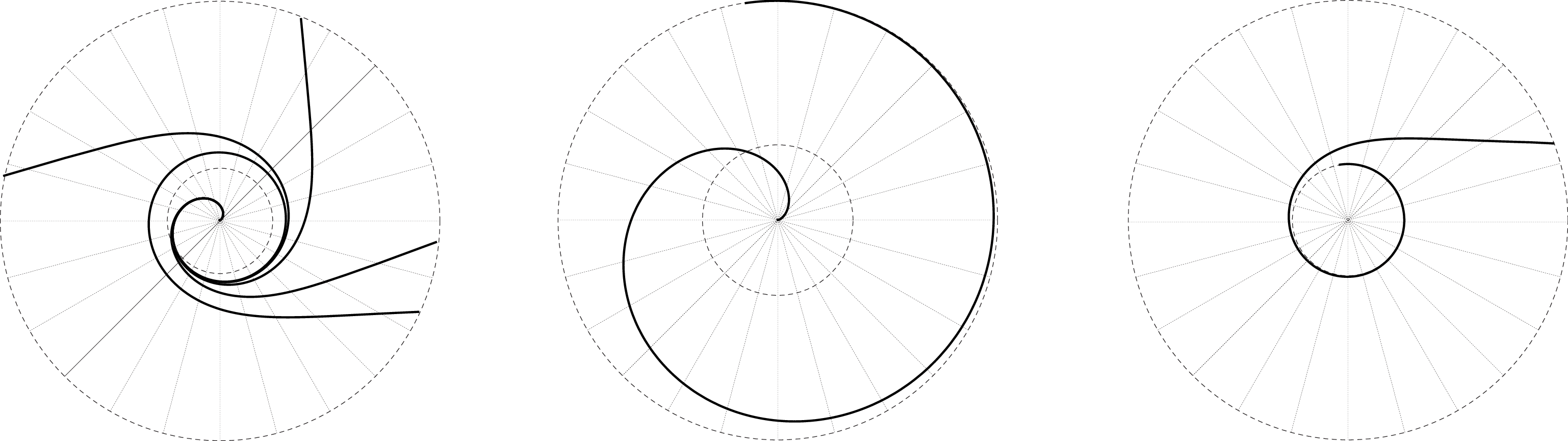}
	\caption{Plots referring to Eq. (\ref{polar1}) in a cut of the $\phi=\pi/2$-invariant plane of the non-compact hyperbolic topological Weyl black hole spacetime. LEFT PANEL: Unbounded trajectory for massless particles with $b<b_c$. The fall to the cosmological horizon or event horizon depends on the initial direction of their motion. MIDDLE PANEL: Critical trajectory of the first kind, in which $r_i<r_m$ and $b=b_c$. Photons can asymptotically fall to a circle of radius $r_m$ or fall to the event horizon. RIGHT PANEL: Critical trajectory of the second kind, in which $r_i>r_m$ and $b=b_c$. Photons can asymptotically fall to a circle of radius $r_m$ or fall to the cosmological horizon.}
	\label{f3} 
\end{figure}

In order to obtain the analytical expression for the orbit in which the photons are moving, we perform an integration of Eq. (\ref{radialpolareq}), resulting in
\begin{eqnarray}\label{polar1}
r(\phi)&=&|x_1|\pm{1\over 4\wp(\kappa\,\phi;\, g_{2},g_{3})-\alpha_1 /3},\qquad {\rm if}\quad b_{c} <b<\infty,\\ \label{polar2}
r(\phi)&=&{1\over 4\wp(\kappa\,\phi+\omega_0;\, g_{2},g_{3})-\alpha_1 /3},\qquad {\rm if}\quad 0<b\leq b_{c},
\end{eqnarray}
where $\wp(x; g_2, g_3)$ is the $\wp$-Weierstra{\ss} elliptic function, whose Weierstra{\ss} invariants are given by
\begin{equation}\label{invhyp}
g_2 ={1 \over 4}\left({\alpha_{1}^{2} \over 3}-
\beta_1^2\right),\qquad
g_3 ={1 \over 16}\left({\alpha_{1}\beta_1^2 \over 3}-{2 \over 27}\alpha_{1}^{3}-\gamma_1^3\right),
\end{equation}
whereas the constants appearing in Eqs. (\ref{polar1}-\ref{invhyp}) are given explicitly by
\begin{eqnarray}\nonumber
\kappa&=&\frac{\sqrt{x_1\,x_2\,x_3}}{\mathcal{B}},\qquad
\alpha_1={1\over x_1}+{1\over x_2}+
{1\over x_3}\\ \nonumber
\beta_1&=&\left[{1\over x_1\,x_2}+
{1\over  x_1\,x_3}+
{1\over x_2\,x_3}\right]^{1/2},\qquad
\gamma_1=\frac{1}{[x_1\,x_2\,x_3]^{1/3}}.
\end{eqnarray}	Also, the plus and minus sign in Eq. (\ref{polar1}) correspond to orbits of the first kind ($r \geq r_d$) and orbits of the second kind ($r \leq r_a$), respectively (see right panel of Fig. \ref{f2}), and the values of $x_j$ ($j = 1, 2, 3$) depend on the region where the motion occurs as well as the type of orbit. Table \ref{table1} summarizes the values used in Equations (\ref{polar1}) and (\ref{polar2}), which depend on the value of the impact parameter, whereas in Fig. \ref{f3} the trajectories obtained from Equation (\ref{polar2}) are shown, i.e., for photons with $b\leq b_c$.
\begin{table}[h!]
\begin{center}
		\label{table1}\caption{Quantities used in Eqs. (\ref{polar1}) and (\ref{polar2}) to obtain the trajectories of the photons in the exterior geometry of a non-compact hyperbolic topological black hole spacetime in conformal Weyl gravity.}
	\begin{tabular}{|c|c|c|c|c|c|c|}
		\hline
Type of orbit & Range for $r$ & Range for $b$ & $x_1$ & $x_2$ & $x_3$ & $\omega$ \\\hline
{\it First kind}	& $r>r_d$ & $b_c<b<\infty$ & $r_d$ & $r_d-r_a$ & $r_d-r_1$ & -- \\ \hline
{\it Second kind}	& $r<r_a$ & $b_c<b<\infty$ & $-r_a$ & $r_d-r_a$ & $-(r_a-r_1)$ & -- \\ \hline
{\it Critical FK} & $r>r_m$ & $ b = b_c$ & $r_1$ & $-r_d$ & $-r_a$ & $\wp^{-1}(\frac{\alpha_1}{12})$ \\ \hline
{\it Critical SK} & $r<r_m$ & $b = b_c$ & $r_1$ & $-r_d$ & $-r_a$ & 0 \\ \hline
{\it Unbounded}	& $r>r_+$ & $0<b< b_c$ & $r_1$ & $-r_d$ & $-r_a$ & 0 \\ \hline 
	\end{tabular}
\end{center}
\end{table}

\subsection{Bending of light}
As an application of the preceding section, specifically of the polar equation (\ref{polar1}) for the photon orbit of the first kind, it is possible to study how this is deflected due to presence of the non-compact black hole, whose singularity, for the sake of simplicity, is placed at the origin of a coordinate system $O$. Therefore, in order to obtain a full description, we employ the exact method outlined by Villanueva \& Olivares \cite{vo13} instead of the Rindler--Ishak method performed in other works \cite{Bha09,RI07,RI10,IRD10,FGG07}.

\begin{figure}[!h]
	\begin{center}
		\includegraphics[width=120mm]{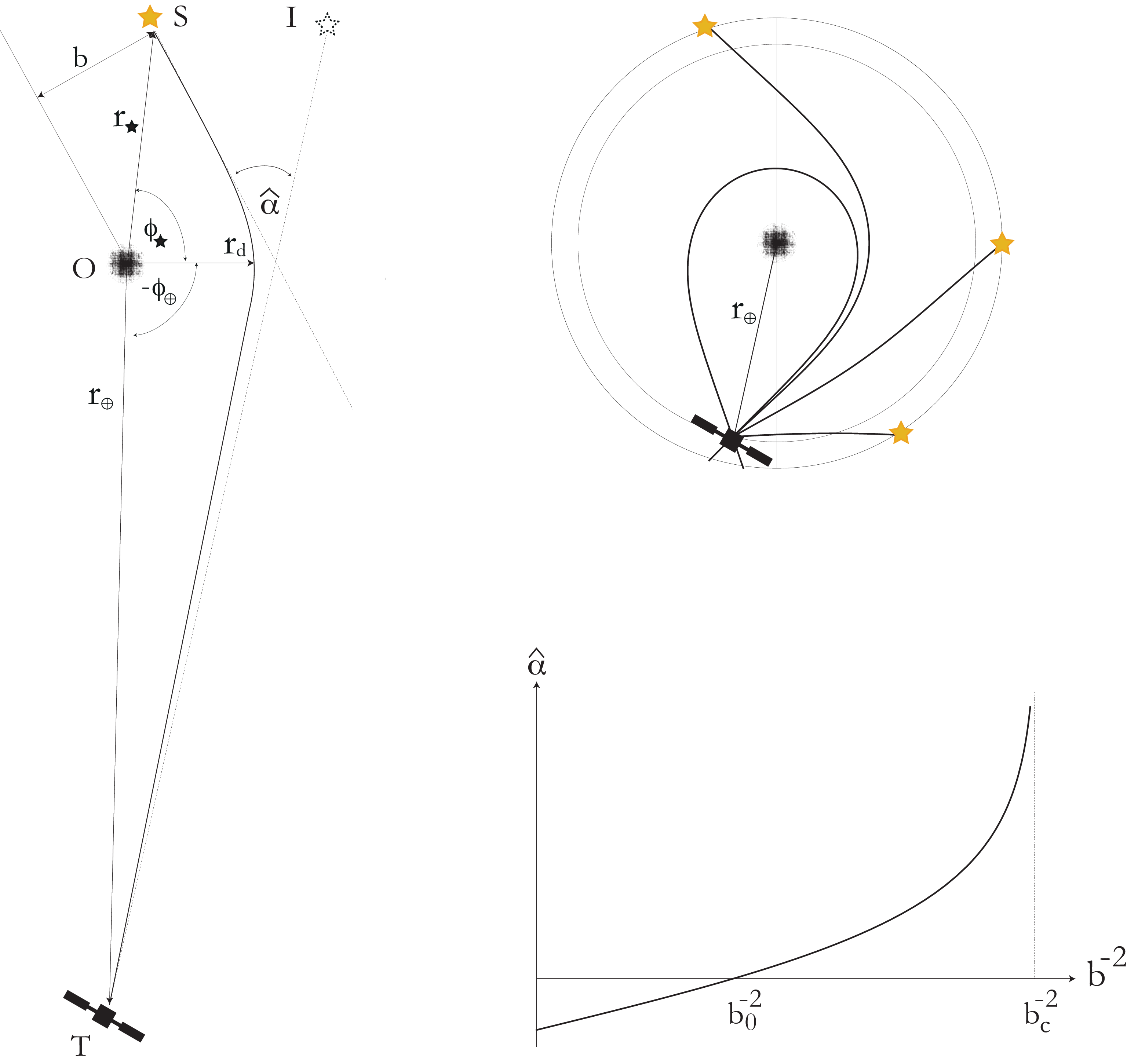}
	\end{center}
	\caption{Gravitational lensing from a non-compact hyperbolic topological black hole in conformal Weyl gravity.
		LEFT: Light with impact parameter $b$ from the source $S$ at $(r_{\bigstar}, \theta_{\bigstar})$ bends at the lens $O$ and arrives at the observer $T$ at $(r_{_{\bigoplus}}, -\theta_{_{\bigoplus}})$,  who then sees the image $I$; TOP RIGHT: Trajectories for different values of the impact parameter as observed at $T$;  BOTTOM RIGHT: The angle of deflection $\widehat{\alpha}$ as a function of $b^{-2}$. There is a critical value of the impact parameter $b_0$, for which the deflection angle vanishes and thus the bending of light can be attractive or repulsive. Also, the deflection angle approaches infinity as the impact parameter approaches the critical value $b_c$.}
	\label{f4}
\end{figure}

Suppose that in $t=t_0$ photons are emitted with impact parameter $b_{c} <b <\infty$ by a source placed at $(r_i, \theta_i)=(r_{\bigstar}, \theta_{\bigstar})$ with respect to $O$. Therefore, photons travelling in the exterior spacetime of the non-compact hyperbolic topological black hole are deflected in such way they reach the turning point $r_d$ given by Eq. (\ref{distapelio}), and then are received by an observer placed at the position $(r_{\bigoplus}, -\theta_{\bigoplus})$ on the manifold (see left-top panel of Fig. \ref{f4}). Based on the same plot, it is not difficult to prove that \begin{equation}\label{defangdefl}
	\widehat{\alpha}=|\theta_{\bigstar}|+|\theta_{_{\bigoplus}}|-\pi;
\end{equation}
therefore, by combining Eq. (\ref{polar1}) with Eq. (\ref{defangdefl}), the deflection angle becomes 
\begin{equation}\label{angdef}
\widehat{\alpha} = {1\over\kappa}\left(
\left|\wp^{-1}\left[\frac{1}{4( r_{\bigstar}-r_d)}+\frac{\alpha_1}{12}\right]\right|+
\left|\wp^{-1}\left[\frac{1}{4( r_{\bigoplus}-r_d)}+\frac{\alpha_1}{12}\right]\right|\right)-\pi,
\end{equation}
where $\wp^{-1}$ is the inverse $\wp$-Weierstra{\ss} function \cite{byrd,hancock,Armitage,tablas}. In the bottom left panel of Fig. \ref{f4}, the deflection angle as a function of $b^{-2}$ is plotted, and it shows that there is a special value at which the deflection angle vanishes. This means that the scattering of photons by a hyperbolic topological black hole can be attractive or repulsive (with respect to the black hole), depending on the influence of the event horizon or the cosmological horizon (see top and bottom right panel of Fig. \ref{f4}).

\section{Summary}

In this research, we have studied the null geodesics of a non-compact hyperbolic topological black hole spacetime in conformal Weyl gravity, the metric of which is given by Eq. (\ref{lapshyp}). In this context, we analyze the existence of the event horizon, $r_+$, and the cosmological horizon, $r_{++}$, as a function of the cosmological parameter $k$, and the Weyl parameter, $\gamma$, as shown in Fig. \ref{graf1}, and we obtain their analytical expression in terms of these parameters. Also, the Killing vector fields are obtained and the associated four conserved quantities, showing the stationarity and the rotational invariance of the metric. These symmetries reduce our study to a unidimensional equivalent problem on an invariant plane, which is chosen to  be $\phi=\pi/2$.
A quick view of the equations of motion show us that the radial motion of photons presents the same features performed in the standard conformal Weyl gravity \cite{vo13}, and therefore, in the same way as Einstein's black holes, photons cross the horizons in a finite proper time, whereas in an external system, the photons fall asymptotically to the horizons.

The angular motion was first studied qualitatively by means of an analysis of the effective potential as a function of the radial distance for different values of the impact parameter $b$ (see Fig. \ref{f2}). Therefore, under the condition $\gamma < \gamma_c$ and $k < k_{ext}$, this analysis shows the existence of a maximum for which photons with $b = b_c=(k_{ext}-k)^{-1/2}$ either fall to one of the horizons or tend asymptotically to an unstable circular orbit at $r_m$. Also, we note that the existence of the unstable circular orbit depends on the parameter $\gamma$, which is a novel result because in the standard  conformal Weyl gravity this only depends on the mass $M$ \cite{vo13}. For this reason, the proper period also depends on the Weyl parameter $\gamma$, whereas the coordinate period, as we expect, depends on $\gamma$ and $k$. However, if the value of the impact parameter is $0<b<b_c$, then photons fall inexorably to one of the horizons. Finally, when the impact parameter is $b=b_d$ so $b_c<b_d<\infty$, the trajectory is deflected in such a way that for $r\leq r_a$ then $r_a$ corresponds to an apoastron distance, or for $r\geq r_d$ the distance $r_d$ corresponds to a periastron distance. The exact calculation of both distances, $r_a$ and $r_d$, makes it possible to verify that in the limit $b\rightarrow \infty$, the consistency relations $r_a\rightarrow r_+$ and $r_d \rightarrow r_{++}$ are satisfied.

For the quantitative analysis of the motion, we employ an integration of the general equation of motion (\ref{angeqmot}) in such way that we obtain two general equations for the trajectories described for the photons: unbounded and critical trajectories ($0<b\leq b_c$) are described by Eq. (\ref{polar1}) and are shown in Fig. \ref{f3}, whereas orbits of the first and the second kind are described by Eq.(\ref{polar2}), in which case the orbits are bounded.  

As an application of our previous results, we obtain an exact expression in terms of the inverse $\wp$-Weierstra{\ss} function, for the deflection angle $\widehat{\alpha}$ experienced by photons emitted with impact parameter $b_d$ by a source with coordinates ($r_{\bigstar}, \theta_{\bigstar}$) with respect to the black hole system (which acts as a gravitational lens). The photons are then received by an observer at ($r_{\bigoplus}, -\theta_{\bigoplus}$) (see left top panel of Fig. \ref{f4}). One important and interesting result is the physical type of the deflection experienced by the massless particles. Depending on the value of the impact parameters this can be either attractive or repulsive with respect to the black hole. This feature is better appreciated in the bottom left panel of Fig. \ref{f4}, in which the deflection angle $\widehat{\alpha}$ vanishes at a specific value of the impact parameter $b_0$.

\begin{acknowledgments}
This work was funded by the Comisi\'on Nacional de Investigaci\'on 
Cient\'ifica y Tecnol\'ogica through FONDECYT Grant No. 1181708 (GC), PAI Grant No. 79150053 (GC) and DOCTORADO Grant No. 21181476 (CO).

\end{acknowledgments}

\end{document}